\documentclass[aip,rsi,preprint,graphicx]{revtex4-1} 

\usepackage{graphicx} 
\usepackage{dcolumn} 
\usepackage{bm} 
\usepackage{mathtools,amsmath,amssymb}
\usepackage{hyperref}
\usepackage{url}
\usepackage{multirow}
\usepackage[utf8]{inputenc}

\draft 

\usepackage[T1]{fontenc}
\usepackage{mathptmx}
\usepackage{etoolbox}

\makeatletter
\def\@email#1#2{%
	\endgroup
	\patchcmd{\titleblock@produce}
	{\frontmatter@RRAPformat}
	{\frontmatter@RRAPformat{\produce@RRAP{*#1\href{mailto:#2}{#2}}}\frontmatter@RRAPformat}
	{}{}
}%
\makeatother

\begin{document}


\title{On the origin and elimination of cross coupling between tunneling current and excitation in scanning probe experiments that utilize the qPlus sensor 
\footnote{The following article has been submitted to Review of Scientific Instruments on March 23, 2023, but has not yet been published.}
} 



\author{Michael Schelchshorn}
 \email{michael.schelchshorn@ur.de}
\author{Fabian Stilp}
\author{Marco Weiss}
\author{Franz J. Giessibl}
 \email{franz.giessibl@ur.de}
\affiliation{University of Regensburg, Institute of Experimental and Applied Physics, Universit\"atsstrasse 31, D-93040 Regensburg, Germany.}


\date{\today}

\begin{abstract} 
The qPlus sensor allows simultaneous measurements of scanning tunneling microscopy (STM) and atomic force microscopy (AFM). 
Its design for use in frequency modulation AFM (FM-AFM) involves separate electrodes, applied on an oscillating quartz cantilever, for the detection of an electronic tunneling current and the deflection signal. Cable resistance and capacities in the electronic setup can induce cross talk phenomena. 
We report a tunneling current-induced cross coupling observed in a combined STM/AFM setup which uses the qPlus sensor. This cross coupling can induce a positive or negative change of the amplitude excitation signal, thus acting as an apparent dissipation or drive. The experimental data is explained well by tunneling current-induced fluctuations of the electric potential on the sensor electrodes.
\end{abstract}


\maketitle 


\section{Introduction\label{chpt_introduction}}

Scanning tunneling microscopy (STM) \cite{Binnig1982b,Binnig1982,Binnig1983} and atomic force microscopy (AFM) \cite{Binnig1986, BinnigAFMPatent1986} allow for mapping, identification, and manipulation of individual atoms and molecules on various surfaces and samples. \cite{Chen2021,Giessibl1992UM,Morita2002,Morita2009,Morita2015,Garcia2002SurfSciRep,Giessibl2003RMP,Giessibl2019RSI} While the operation of STM depends on a current flow between tip and sample and is thus limited to conductive tips and surfaces, AFM utilizes the forces between atoms of the tip and the surface for image generation. \\
In addition to the technical challenges of STM, however, AFM faces extra experimental difficulties, e.g., a non-monotonic force-distance behavior, long-range contributions due to Van-der-Waals interactions, and a jump-to-contact phenomenon; \cite{Giessibl2003RMP} atomic resolution with AFM was thus not achieved until 1992, \cite{Giessibl1992UM,Binnig1992UM,Ohnesorge1993} six years after the method was originally introduced. The benchmark test for atomic resolution, resolving the 7x7 reconstruction of the Si(111) surface, had to wait another three years, \cite{Giessibl1995} and ultimately became possible by employing frequency modulation AFM (FM-AFM), a method that was introduced four years earlier to resolve micrometer sized magnetization patterns in a fast and sensitive non-contact imaging mode. \cite{Albrecht1991} \\

Quartz tuning forks as used in wrist watches have been used in scanning probe microscopy for a long time for instance in acoustic near field microscopy, \cite{Guethner1989} for distance control in scanning near field optical microscopy \cite{Karrai1995} and for cryogenic scanning force microscopy. \cite{Rychen1999}
The qPlus sensor is a stiff quartz cantilever that was originally also built from a quartz tuning fork, \cite{Giessibl1998APL} with one of the prongs being attached to a heavy substrate, while the other prong was fitted with a probe tip. As atomic resolution was soon obtained after its invention, \cite{Giessibl2000APL} its popularity grew rapidly and it is now used in hundreds of instruments all over the world. \cite{Giessibl2019RSI} \\ 
With its small oscillation amplitudes and possibility to use metallic tips, the qPlus sensor facilitates simultaneous STM and AFM in one instrument. However, when measuring two signals, like the tunneling current and the signal resulting from the forces at the tip, at the same time, a cross coupling can occur. This effect has been observed since the early years of combined STM/AFM experiments using qPlus sensors; a typical solution is suggested by Heyde et al. \cite{Heyde2006} who mount the tip using insulating epoxy to the qPlus sensor and provide an extra wire that carries the tunneling current or bias voltage. Although this method prevents cross coupling effectively, the mechanic effect of the wire often leads to reduced $Q$ factors and multiple resonances. \\

Regarding an explanation of the origin of cross coupling different examples have previously been discussed in the literature. Weymouth et al. \cite{Weymouth2011PRL} report a cross talk between current and force leading to a repulsive \lq\lq phantom\rq\rq{} force on samples with limited conductivity like semiconductors. This phantom force leads to a reduction in the attractive electrostatic force. Majzik et al. \cite{Majzik2012} analyze non-idealities of current-to-voltage converters like the STM pre-amplifier and virtual ground issues (the oscillation of the virtual ground potential of the STM pre-amplifier) as a possible reason for cross talk between tunneling current and deflection measurement: an unwanted capacitive coupling of the STM virtual ground to the input of the AFM pre-amplifier changes the deflection signal obtained by the latter; they present an improved sensor design (similar to Refs.~\onlinecite{Heyde2006,Albers2008}) to lower the coupling of the STM and AFM channels. Replacing the internal coaxial cable with a double-shielded one, reduces the stray capacitance further.
Cross coupling in the case of a sensor design with the tip contacted via an extra wire is examined by Nony et al. \cite{Nony2016} who suggest an electromagnetic radiation-induced coupling between the oscillation of the quartz sensor beam and the tunneling current flowing through that wire. \\

In this study, we report on a cross coupling that inflicts a change of the excitation amplitude of the sensor in our setup, depending on the sign of the tunneling current. At one bias polarity, the cross coupling acts as an excitation of the oscillation, in the opposite bias it appears as a damping. The root cause of this current-induced apparent dissipation is that while the piezoelectric effect is used to generate an electrical current from the vibration, the opposite is also true. A modulation of the voltage on the electrodes of the sensor will induce a deflection. Just like microphones can be used as loudspeakers, sensors often can serve as actuators and vice versa. Such a back-action of the deflection detection on the oscillator is often observed, in optical setups as well as in piezoresistive detection. \cite{Giessibl1997APL} \\
We find that our qPlus sensor bends by approximately 180\,pm when the differential voltage between the STM electrode and the deflection electrodes (see Fig.~\ref{fig_qPlus}) changes by 1\,V (see section~\ref{sec_alphabeta}). Thus, an AC voltage at the sensors resonance frequency of that tip electrode merely needs an amplitude of 1\,$\mu$V to excite the sensor to an amplitude of about 36\,pm if the quality factor of the sensor is $Q=200\,000$ and we later used that exact electrode for the excitation of the oscillation. If the tunneling current that flows through one of the sensor electrodes modulates its electrical potential by 1\,$\mu$V at a phase of $90^\circ$ (at resonance), a change of the excitation signal of $\pm 72$\% will occur for an amplitude setpoint of 50\,pm. In theory, the tunneling current has a phase shift of $180^\circ$ with respect to the sensor deflection and therefore should not alter the excitation signal. In practice, cable resistance and parasitic or intentional capacities introduce RC circuits into the current line that cause an additional phase shift, resulting in unwanted damping or excitation. \\
We present a model that explains the cross coupling between tunneling current and sensor excitation, supported by experiments and supplemented by verified measures on how to prevent or at least minimize cross talk. 

\section{General concepts \label{chpt_basic}}

\subsection{STM \label{sec_STM}}
 
Scanning tunneling microscopy (STM) is based on the detection of a tunneling current between a sharp, metallic probe tip and a conductive sample, the surface of which is scanned at a distance of a few 100\,pm. If a bias voltage $V_{\text{B}}$ is applied between tip and sample, a quantum-mechanical tunneling current flows, \cite{Chen2021} which depends exponentially on the distance $z$ between tip apex and sample surface: \cite{Binnig1982}
\begin{equation}
	I_{\text{t}}(z)=I_0\times\exp(-2\kappa z), \label{eq_tunneling_current}
\end{equation}
where $I_0 = I_{\text{t}} (z=0)$ is determined by the bias and $\kappa$ is the decay constant, which depends on the work functions of tip and sample, respectively. 

\subsection{FM-AFM \label{sec_FM_AFM}}

In frequency modulation atomic force microscopy (FM-AFM) the quartz cantilever is driven at resonance while fixing the oscillation amplitude $A$ by feeding the 90 degree phase-shifted deflection signal through a Proportional-Integral (PI)-amplitude controller back to the cantilever. When the oscillating tip is subject to a nonzero tip-sample force gradient $k_\text{ts}$, the oscillation frequency changes from the fundamental eigenfrequency of the free cantilever $f_0$ by the frequency shift $\Delta f$, \cite{Giessibl1997PRB} which is given by $f_0\times\langle k_{\text{ts}}(z_0) \rangle/(2k)$ where the pointed brackets indicate an averaging process as explained in Ref.~\onlinecite{Giessibl2003RMP}. \\

Another experimental observable of FM-AFM with oscillating cantilevers like qPlus is the dissipative component of the tip-sample interaction. \cite{Denk1991APL} 
Dissipation data was explained using the Tomlinson-Prandtl model \cite{Prandtl1928,Tomlinson1929,Cleveland1998APL,Sasaki2000} of friction as a plucking action on single atoms. \\ 
The energy dissipated over one oscillation cycle $\Delta E_{\text{ts}}$ is given by \cite{Sasaki2000}
\begin{equation}
	\Delta E_{\text{ts}} = \oint \vec{F}_{\text{ts}} \,\text{d}\vec{z} = -\int_0^{2\pi} F_{\text{ts}}(z+A\cos{\phi})A\sin{\phi} \,\text{d}\phi, \label{eq_dE_gen}
\end{equation}
which is nonzero if a hysteresis occurs in the tip-sample force curve  $F_{\text{ts}}(z)$ over the $z$-range covered by the oscillating cantilever.
In order to keep the amplitude $A$ constant, the drive signal $X'_{\text{drive}}$ changes as follows: \cite{Giessibl2003RMP}
\begin{equation}
	\frac{X'_{\text{drive}}}{X_{\text{drive}}} = 1 + \frac{Q}{2\pi}\frac{\Delta E_\text{ts}}{E}, \label{eq_dE}
\end{equation} 
where $X_{\text{drive}}$ is the amplitude of the excitation signal of the free oscillation (with frequency $f=f_{0}$ and a phase shift $\phi_\text{drive}=90^\circ$ with respect to the sensor oscillation); internal dissipation in the (free) motion of the cantilever itself is defined by its quality factor $Q$ and the total oscillation energy for a cantilever with stiffness $k$ and oscillation amplitude $A$ is given by $E = \frac{1}{2}\,k\,A^2$.

\subsection{Dynamic STM\label{sec_tunneling_current}}

\begin{figure}
	\centering
	\includegraphics{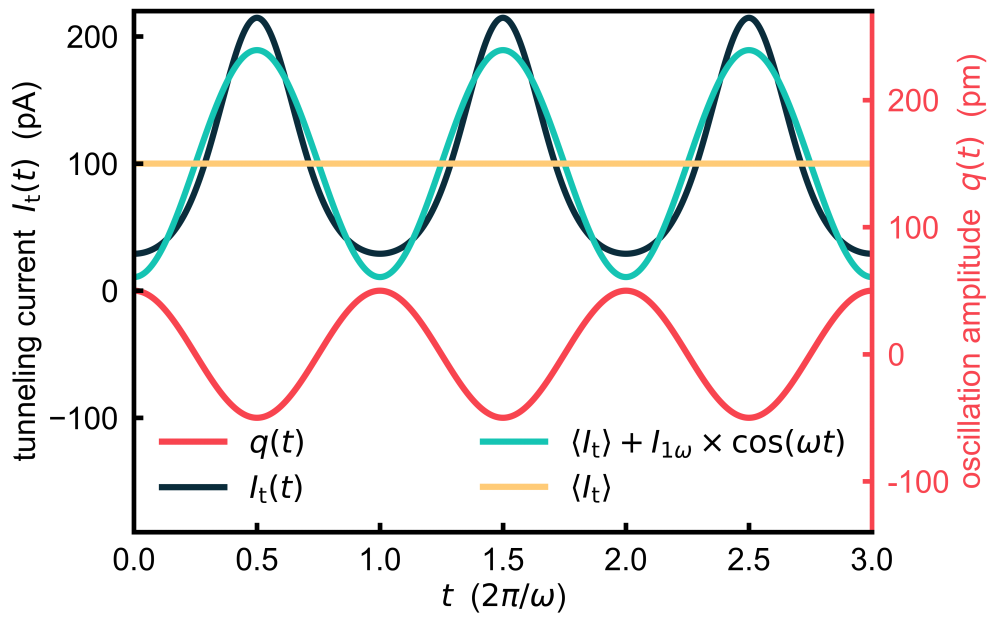}
	\caption[The tunneling current $I_{\text{t}}(t)$ and its Fourier expansion $I_{\text{t}}$ as a function of time.]{\label{fig_I_t_plot} The tunneling current $I_{\text{t}}(t)$ as a function of time computed according to Eq.~\eqref{eq_tunneling_current_osc}, the time-averaged current $\langle I_{\text{t}} \rangle$ as well as the Fourier series (Eq.~\eqref{eq_I_t_fourier}) up to first order in $\omega$. The oscillation amplitude $q(t)=A\times\cos(\omega t)$ is plotted for reference. The parameters used for the plots are $A=50\,\text{pm}$, $\kappa = 1\times10^{10}\,\text{m}^{-1}$, and $\langle I_{\text{t}} \rangle = 100 \,\text{pA}$.}
\end{figure}

When combining AFM and STM measurements by FM-AFM, the electronic tunneling current between tip and sample varies with time due to the sensor's oscillation. Assuming a harmonic tip motion with amplitude $A$ and angular frequency $\omega = 2\pi\times f = \sqrt{(k+k_{\text{ts}}(z_0))/m^*}$, without higher harmonics, the instantaneous tip-sample separation can be written as:
\begin{equation}
	z(t)=z_0+A\times\cos(\omega t), \label{eq_z(t)}
\end{equation}
where $z_0$ is the tip-sample distance at rest, which is modulated by the cantilever motion $q(t) = A\times\cos(\omega t)$. In accordance with Eq.~\eqref{eq_tunneling_current}, we obtain the tunneling current as a function of time $t$ (see Fig.~\ref{fig_I_t_plot}):
\begin{equation}
	I_{\text{t}}(t)= I_{\text{z}_0}\times\exp\big(-2\kappa A\times\cos(\omega t)\big) \label{eq_tunneling_current_osc}
\end{equation}
with $I_{\text{z}_0} = I_{0}\times\exp(-2\kappa z_0)$. As $I_{\text{t}}(t)$ is periodic in time, it can be expanded in a Fourier series: 
\begin{equation}
	I_{\text{t}}(t) = \langle I_{\text{t}}\rangle + I_{1\omega}\times\cos(\omega t) + I_{2\omega}\times\cos(2\omega t) + ..., \label{eq_I_t_fourier}
\end{equation}
where the second order term $I_{2\omega}\times\cos(2\omega t)$ and all higher terms are small compared to the first order term if the oscillation amplitude $A$ is small compared to the inverse decay rate of the tunneling current $1/\kappa$. The DC and first order terms are given by \cite{Huber2015}
\begin{equation}
	\langle I_{\text{t}}\rangle = I_{\text{z}_0}\,\mathfrak{I}_0(2\kappa A) \quad \text{and} \quad I_{1\omega}=-2\,\langle I_{\text{t}}\rangle\,\frac{\mathfrak{I}_1(2\kappa A)}{\mathfrak{I}_0(2\kappa A)}, \label{eq_I_t_Fourier_coeff}
\end{equation}
respectively, and $\mathfrak{I}_n$ is the modified Bessel function of first kind of order $n$. $\langle I_{\text{t}}\rangle$ is equal to the time-average of $I_{\text{t}}(t)$ over one oscillation period; it denotes the measured value of the tunneling current in experiment as the bandwidth of the tunneling current amplifier is typically small compared to the oscillation frequency. \cite{Huber2013} When using a tunneling current amplifier with a high bandwidth, the first order term can be used to measure the decay rate and thus the work function at atomic resolution. \cite{Herz2005} \\

\section{Experimental setup and first observation of apparent dissipation \label{sec_setup_observation}}

\subsection{Experimental setup \label{sec_exp_setup}}

\begin{figure}
	\centering
	\includegraphics{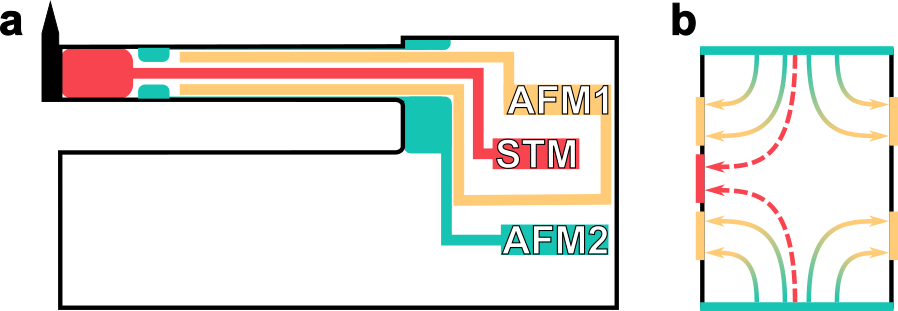}
	\caption[Geometry of the qPlus sensor in our microscope.]{Geometry of the qPlus sensor in our microscope. \textbf{a} Electrode layout. Two AFM contacts are used for differential deflection measurement and one STM contact for detecting the tunneling current. \textbf{b} Cross-section through the oscillating beam in \textbf{a}. The electric field distribution caused by a deflection of the sensor (solid arrows) is shown. This phenomenon is known as the piezoelectric effect, the inverse of which is also possible: a modulation of the potential $V_{\text{s}}$ on the STM electrode leads to a deflection of 180\,pm/V.
		\label{fig_qPlus}}
\end{figure}

All measurements shown here have been conducted with a home-build low-temperature ($T\approx 5.5\, \text{K}$) combined scanning tunneling and atomic force microscope operating in ultra-high vacuum ($p\approx 5\times10^{-11}\,\text{mbar}$). \cite{Emmrich2015, Schneiderbauer2014}\\		
The sensor used for combined STM- and AFM-measurements in our microscope is based on a third-generation qPlus sensor (Type S1.0 in Table 1 in Ref.~\onlinecite{Giessibl2019RSI}) with a stiffness of $k=1800\,\text{N}/\text{m}$. An etched tungsten tip is attached to its oscillating prong and an electrode dedicated to collecting the tunneling current is applied on one side of the oscillating quartz beam as shown in Fig.~\ref{fig_qPlus}. Additional electrodes are located on each side of the beam, whereby electrodes on opposite sides are electrically connected. The two corresponding AFM contacts on the sensor serve as differential input for the AFM pre-amplifier. Our sensor is driven by connecting the amplitude drive signal to the $z$-terminal of the xyz scan piezo to shake the sensor holder. The amplitude of this AC voltage serves as the excitation signal $X'_{\text{drive}}$ in the measurement: a positive amplitude corresponds to a driving, a negative value to a damping force. \\
The resonance frequency $f_0$ and quality factor $Q$ are obtained from a frequency sweep: $f_0 = 20419.3 \,\text{Hz}$ and $Q = 210320$. \\

\begin{figure}
	\centering
	\includegraphics{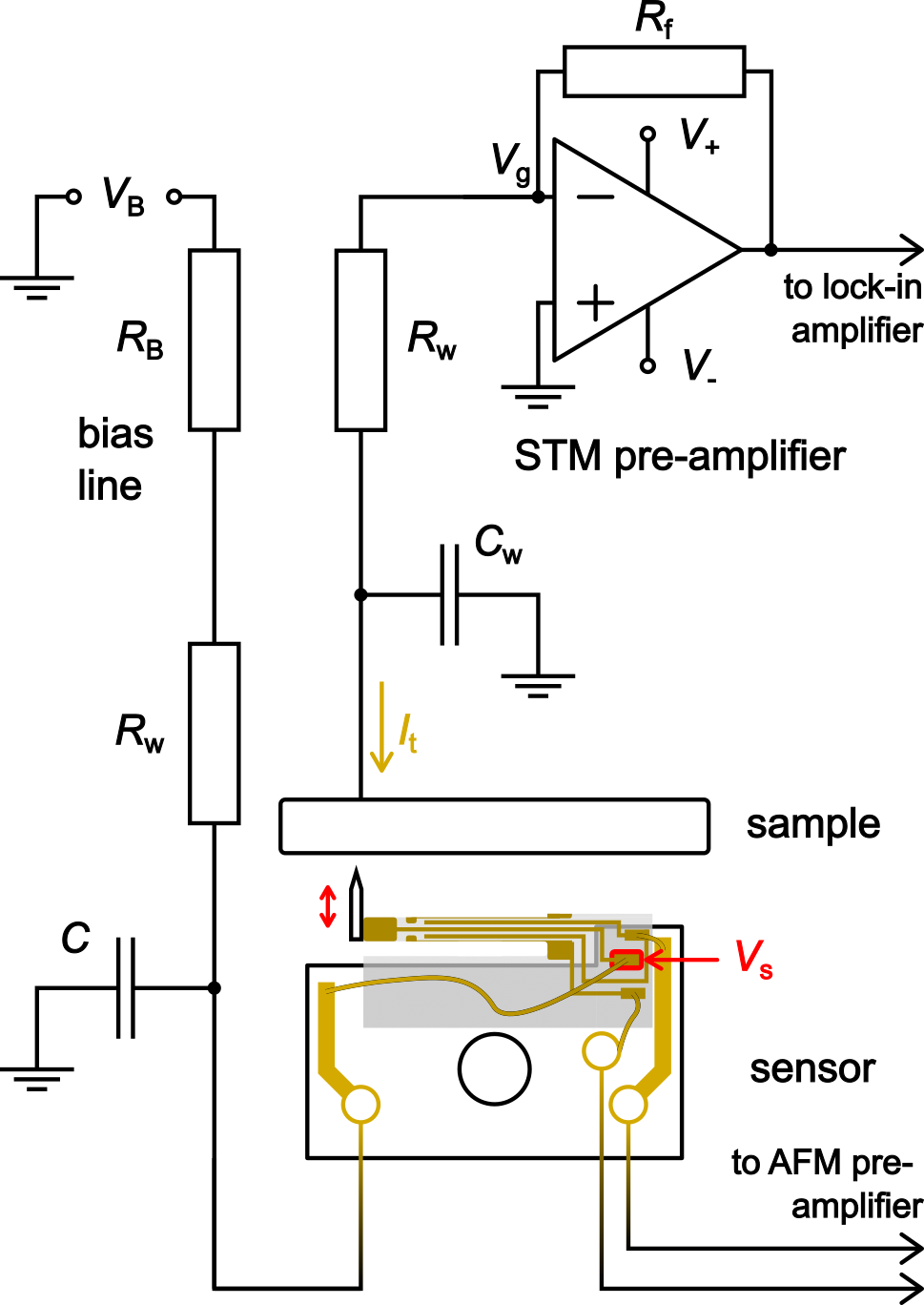}
	\caption{Relevant section of the microscope's STM electronics. The bias is applied at the tip and the sample connects to the inverting input of the STM pre-amplifier, the common reference potential is the ground of the microscope chamber. The positive direction of the tunneling current, i.e., the (technical) current direction for negative bias voltage $V_{\text{B}}$, is indicated by an arrow; thus $I_\text{t}(t)$ and $V_\text{s}(t)$ have opposite sign.
		\label{fig_STM_electronics}}
\end{figure}

The origin of cross coupling here, between the excitation signal and the tunneling current, is directly related to the wiring of the qPlus sensor and the sample for simultaneous detection of tunneling current and sensor deflection. The relevant section of the electronics is shown in Fig.~\ref{fig_STM_electronics}. For a broader description of the microscope's setup, we refer to Refs.~\onlinecite{Emmrich2015, Schneiderbauer2014, Huber2018, Berwanger2019}.\\
In the configuration of Fig.~\ref{fig_STM_electronics}, the tip of the qPlus sensor is biased. The line between the bias voltage output and the STM contact of the sensor includes an external resistor $R_\text{B}=110\,\text{k}\Omega$ and a coaxial cable leading to the microscope, which has negligible resistance. Inside the microscope, high-resistance coaxial cables are used to prevent thermal coupling with the outside: the resistance $R_{\text{w}}$ of such cable is approximately $100\,\Omega$. \cite{Emmrich2015} The combined capacitance $C$ of the coaxial cables to ground (which is typically around $100\,\text{pF}/\text{m}$) represents the capacitance between $V_{\text{s}}(t)$ and ground potential since the capacitance of the tip-sample junction is on the order of a few picofarad and thus small enough to be neglected. Using a multimeter, $C = 761\,\text{pF}$ was measured. \\

The resistance $R = R_\text{B} + R_\text{w}$ and capacitance $C$ produce a passive low-pass filter to reduce high-frequency noise in the bias voltage signal $V_{\text{B}}$. The transfer function of this RC element is given by
\begin{equation}
	\frac{V_{\text{out}}}{V_{\text{in}}}=\frac{ \frac{1}{i\omega C}}{R+\frac{1}{i\omega C}}=\frac{ 1}{1+ i\omega RC},
	\label{eq_RC_filter}
\end{equation}
where the time constant $\tau=RC$ is inverse to an angular cutoff frequency $\omega_\text{c}=1/\tau$; the output is phase-shifted with respect to the input by $-\arctan(\omega RC)$. For frequencies $f\ll f_\text{c}=\omega_\text{c}/(2\pi) = 1.90\,\text{kHz}$, the output is essentially equal to the input. For $f= f_\text{c}$, the amplitude of $V_{\text{out}}$ is reduced to $V_{\text{in}}/\sqrt{2}$ with a phase shift of $-\pi/4$. For even higher frequencies, the amplitude of $V_{\text{out}}$ is reduced further and the phase shift saturates at $-\pi/2$. \\

The sample is connected to the inverting input of the STM pre-amplifier, which is located just outside the vacuum chamber, by another high-resistance coaxial cable. The STM pre-amplifier is based on a type AD8616, \cite{AnalogDevices} wired as a transimpedance amplifier (see Fig.~\ref{fig_STM_electronics}) with a feedback resistor $R_\text{f}=100\,\text{M}\Omega$. Thus, at a supply voltage of $V_{\pm} = \pm 3\,\text{V}$ the amplifier rails for currents with a magnitude greater than $\frac{3\,\text{V}}{100\,\text{M}\Omega}=30\,\text{nA}$. \\

For an ideal operational amplifier, the input $V_\text{g}$ is always at ground; \cite{Horowitz2015} it is called \lq\lq virtual\rq\rq{} ground. In any real setup non-idealities of the operational amplifier can influence the virtual ground. This effect was characterized by Junk, \cite{Junk2015} who analyzed the modulation of $V_\text{g}$ and its link to cross coupling by considering the influence on the electrostatic force between tip and sample. It was concluded that virtual ground fluctuations can only induce a tiny fraction of the observed apparent dissipation, prompting the present study.\\

To prove experimentally that variations of $V_\text{g}$ can indeed be neglected here, the STM pre-amplifier was disconnected and its port at the microscopy grounded. The same change of the excitation signal as a function of the tunneling current compared to the case with the STM pre-amplifier connected is found, leading to the same apparent dissipated energy in both cases (see appendix~\ref{app_Vg}). Thus, any oscillation of the virtual ground $V_{\text{g}}$ or phase shift in our circuit resulting from the non-idealities of our STM pre-amplifier can be neglected. For all measurements following in the rest of this study, the STM pre-amplifier was connected back to the microscope as depicted in Fig.~\ref{fig_STM_electronics}, while for quantitative considerations and calculations, we assume that $V_{\text{g}}$ is at ground. \\

While non-idealities of the STM pre-amplifier cannot explain the observed signal of the apparent dissipation, the origin of the cross coupling is, instead, linked to the combined effect of an oscillating junction resistance and a high impedance of the bias voltage supply (given by an RC low-pass filter); this RC element leads to an oscillating potential $V_\text{s}(t)$ at the STM electrode of the qPlus sensor, which is phase shifted with respect to the tip oscillation. Due to the piezoelectric nature of the qPlus sensor, this tunneling current-induced modulation of $V_\text{s}(t)$ can couple to the cantilever deflection and thereby lead to an apparent dissipation in measurement. \\
Data for different resistors $R_\text{B}$ shows that the dissipated energy is, in fact, strongly affected when the RC element is changed by varying $R_\text{B}$ (see section~\ref{sec_results} with Fig.~\ref{fig_results}).\\

\subsection{Measurement of apparent dissipation  \label{sec_piezophenom}}

Eq.~\eqref{eq_dE} for the dissipated energy over one oscillation cycle needs to be modified if cross talk between the tunneling current and excitation occurs, replacing $\Delta E_\text{ts}$ by an apparent dissipation $\Delta E_{\text{ts}}^{\text{apparent}}$ with
\begin{equation}
	\Delta E_{\text{ts}}^{\text{apparent}} = \Delta E_{\text{ts}} + \Delta E_{\text{crosstalk}} \label{eq_dE_apparent_gen}
\end{equation}
or
\begin{equation}
	\Delta E_{\text{ts}}^{\text{apparent}} = 2\pi\,\frac{E}{Q}\,\bigg(\frac{X'_{\text{drive}}}{X_{\text{drive}}}-1\bigg). \label{eq_dE_apparent}
\end{equation} 
The experimental manifestation of a change in the driving signal from $X_{\text{drive}}$ (freely oscillating cantilever) to $X'_{\text{drive}}$ (cantilever that oscillates close to the sample) only allows to determine a combined effect of $\Delta E_{\text{ts}}$ and $\Delta E_{\text{crosstalk}}$ as noted in Eq.~\eqref{eq_dE_apparent_gen}. While self-excitation from tip-sample interaction (negative $\Delta E_\text{ts}$), for instance by stochastic motion of hydrogen molecules on Cu(111), \cite{Lotze2012} has been reported, on the bare sample we can expect $\Delta E_\text{ts}$ to be very small, such that the cross talk term dominates the apparent dissipation signal; $\Delta E_{\text{crosstalk}}$ can be positive or negative. \\

To study the apparent dissipation, $z$-spectroscopy measurements were performed over a flat Cu(111) surface: the tip was approached towards the sample in STM feedback beforehand, then retracted 500\,pm from the surface, decreasing the magnitude of the tunneling current $|\langle I_\text{t} \rangle|$ (forward \lq\lq ramping\rq\rq{} direction), and approached back to the starting point, increasing $|\langle I_\text{t} \rangle|$ again (backward \lq\lq ramping\rq\rq{} direction). Due to piezo creep during the measurement the tunneling current at the end can differ from the value at the beginning (see appendix~\ref{app_piezo_creep}). \\
Experiments with $R_\text{B}=110\,\text{k}\Omega$ were performed for the bias voltages $V_\text{B}=\pm0.5\,\text{mV}$, $V_\text{B}=\pm1.0\,\text{mV}$, and $V_\text{B}=\pm10\,\text{mV}$, for nine different amplitude setpoints between $A=10\,\text{pm}$ and $A=400\,\text{pm}$, respectively. Additional measurements with a different resistor, $R_\text{B}=10\,\text{k}\Omega$, and $R_\text{B}=0\,\Omega$ (no external resistor), were carried out with $V_\text{B}=\pm1.0\,\text{mV}$, respectively, for amplitudes between $A=10\,\text{pm}$ and $A=100\,\text{pm}$. These additional measurements are discussed in section~\ref{sec_results}. \\

\begin{figure}
	\centering
	\includegraphics{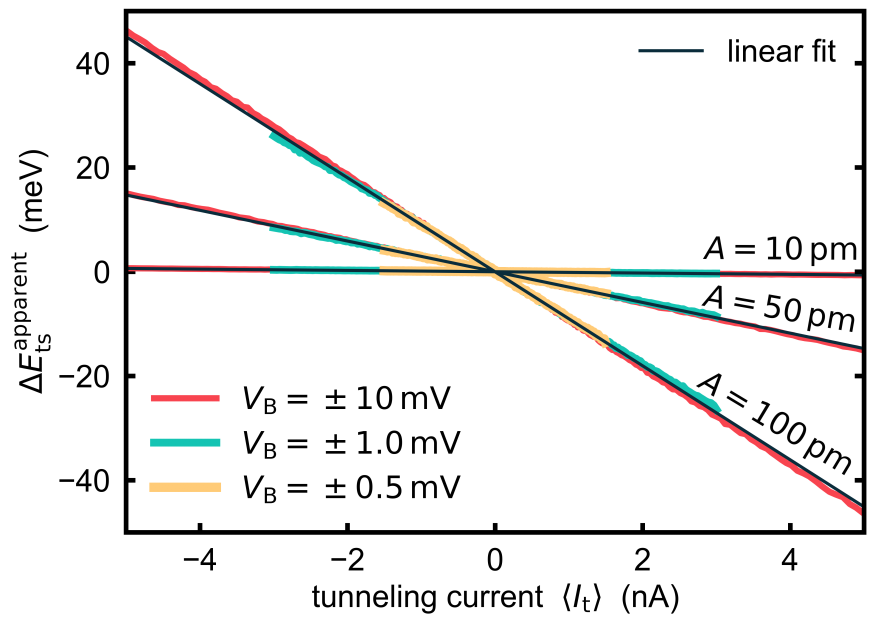}
	\caption{\label{fig_dE} Apparent dissipation per oscillation cycle $\Delta E_\text{ts}^\text{apparent}$ versus tunneling current $\langle I_\text{t}\rangle$ at different bias voltages (forward ramps) and three different amplitudes $A$. All curves intersect at $\langle I_\text{t}\rangle = 0 \,\text{nA}$ and $\Delta E_\text{ts}^\text{apparent} = 0\,\text{meV}$. A linear fit for each of the three amplitudes yields the values listed in Table~\ref{tab_diss}.}
\end{figure}

In Fig.~\ref{fig_dE}, the apparent dissipated energy $\Delta E_{\text{ts}}^{\text{apparent}}$ for $R_\text{B}=110\,\text{k}\Omega$ is plotted versus the tunneling current $\langle I_{\text{t}} \rangle$ for three different amplitudes $A$ by Eq.~\eqref{eq_dE_apparent} with data of excitation $X'_\text{drive}$ from $z$-spectroscopy measurements. The backward direction generally yields the same results as the forward direction (see appendix~\ref{app_piezo_creep}). \\

From the graph (Fig~\ref{fig_dE}), the apparent dissipation indicates three substantial characteristics: \\
1. $\Delta E_{\text{ts}}^{\text{apparent}}$ is linear as a function of the tunneling current $\langle I_{\text{t}} \rangle$. The maximum range of the tunneling current is influenced by the chosen voltage bias $V_\text{B}$ and the tip-sample distance. \\
2. $\Delta E_{\text{ts}}^{\text{apparent}}$ has the opposite sign compared to the tunneling current $\langle I_{\text{t}}\rangle$. For $\langle I_{\text{t}} \rangle<0\,\text{A}$, the oscillation of the tip is damped, while for $\langle I_{\text{t}} \rangle>0\,\text{A}$ the oscillation is driven. At constant current, $\Delta E_{\text{ts}}^{\text{apparent}}$ is independent of the magnitude of $V_{\text{B}}$. \\
3. $\Delta E_{\text{ts}}^{\text{apparent}}$ depends on the amplitude $A$. Larger amplitudes generally provoke a larger apparent dissipation. \\

In summary, for a given amplitude $A$, the apparent dissipation depends only on the tunneling current $\langle I_{\text{t} }\rangle$, whereby the sign of $\langle I_{\text{t}} \rangle$ determines if the apparent dissipation is positive or negative. \\
Values for the apparent dissipated energy per oscillation cycle per tunneling current $\Delta E_{\text{ts}}^{\text{apparent}} \, \big/ \, \langle I_{\text{t}}\rangle$ for the three amplitudes shown in Fig.~\ref{fig_dE} are given in Table~\ref{tab_diss}.

\begin{table}
	\caption{Values for the apparent dissipated energy per oscillation cycle per tunneling current $\Delta E_{\text{ts}}^{\text{apparent}}\, \big/ \, \langle I_{\text{t}}\rangle$ for three different amplitudes $A$, obtained from curves in Fig.~\ref{fig_dE} by a linear fit.
		\label{tab_diss}}
	\begin{ruledtabular}
		\begin{tabular}{cc}
			amplitude $A$  (pm) & $\Delta E_{\text{ts}}^{\text{apparent}} / \langle I_{\text{t}}\rangle$  (meV/nA) \\ 
			
			\hline
			
			$\,\,\,10$ & $-0.13$ \\ 
			
			$\,\,\,50$ & $-2.95$ \\ 
			
			$100$ & $-9.02$ \\
		\end{tabular}
	\end{ruledtabular}
\end{table}

\section{Harmonic model to explain apparent dissipation \label{sec_harmonic_model}}

\subsection{Oscillation of the sensor potential\label{sec_Vs_calculation}}

\begin{figure}
	\centering
	\includegraphics{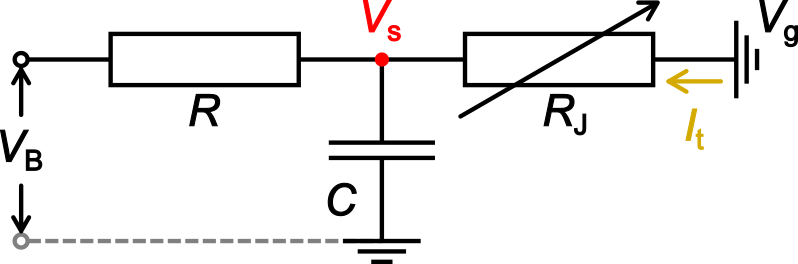}
	\caption[Simplified circuit for the calculation of $V_{\text{s}}(t)$.]{Simplified form of the circuit from Fig.~\ref{fig_STM_electronics} for the calculation of $V_{\text{s}}(t)$. The voltages $V_{\text{B}}$ and $V_{\text{s}}(t)$ are measured against common ground (chamber ground of the microscope); the virtual ground $V_\text{g}$ of the STM pre-amplifier is assumed to be at (common) ground.}
	\label{fig_Vs}
\end{figure}

For the calculation of the potential $V_{\text{s}}(t)$ applied at the STM electrode of our sensor, a simplification of the circuit in Fig.~\ref{fig_STM_electronics} is considered. The resistance $R$ in the bias line and the capacitance $C$ between the STM electrode and chamber ground act as a passive RC element. The tip-sample capacitance is much smaller than the cable capacitance and can be neglected, such that the tip-sample junction can be represented simply by an oscillating resistance $R_\text{J}(t)$. Compared to $R$ and $R_\text{J}(t)$, the line resistance of the cable between sample and STM pre-amplifier is negligible. The inverting input of the latter is assumed to be at ground; $V_\text{g}$ can hence be set directly connected at the sample. The simplified circuit is shown in Fig.~\ref{fig_Vs}.  \\
Kirchhoff's current law for this circuit yields an inhomogeneous linear differential equation for $V_\text{s}(t)$:
\begin{equation}
	\dfrac{V_{\text{B}} - V_{\text{s}}(t)}{R} = C\,\dfrac{\text{d}V_{\text{s}}(t)}{\text{d}t} + \dfrac{V_\text{s}(t)}{R_\text{J}(t)} \label{eq_Kirchhoff}
\end{equation}
or, assuming $V_\text{s}(t) / R_\text{J}(t) = - I_\text{t}(t)$ with $I_\text{t}(t)$ as described in section~\ref{sec_tunneling_current},
\begin{equation}
	\dfrac{\text{d}V_{\text{s}}(t)}{\text{d}t}=-\dfrac{1}{RC}\, V_{\text{s}}(t)+\bigg(\dfrac{I_{\text{t}}(t)}{C}+\dfrac{V_{\text{B}}}{RC}\bigg). \label{eq_V_s_diff_eq}
\end{equation}
The solution of Eq.~\eqref{eq_V_s_diff_eq} for $I_\text{t}(t) = \langle I_\text{t} \rangle + I_{1\omega}\,\cos(\omega t)$ in the steady-state is given by (see appendix~\ref{app_Vs_harm})
\begin{eqnarray}
	V_{\text{s}}(t) &=& V_{\text{B}} + R \, \langle I_{\text{t}} \rangle \nonumber \\
	& &+ \dfrac{R \, I_{1\omega} \, \cos\big(\omega t - \arctan(\omega R C)\big)}{\sqrt{1+(\omega R C)^2}}. \label{eq_V_s_final}
\end{eqnarray}
Thus, given a sinusoidal current, $V_\text{s}(t)$ is also harmonic with the constant offset $V_0 = V_{\text{B}}+R\,\langle I_{\text{t}} \rangle$ plus a cosine with amplitude $V_1 = R\,I_{1\omega} \big/\sqrt{1+(\omega R C)^2}$, phase-shifted with respect to the sensor oscillation by
\begin{equation}
	\phi_{\text{RC}} = -\arctan(\omega R C). \label{eq_phi_RC}
\end{equation}
In our microscope setup, we have $\omega / (2\pi) \approx 20.4 \,\text{kHz}$; thus, $\phi_{\text{RC}} \approx -85^\circ$ for $R=110.1\,\text{k}\Omega$. \\

\begin{figure}
	\centering
	\includegraphics{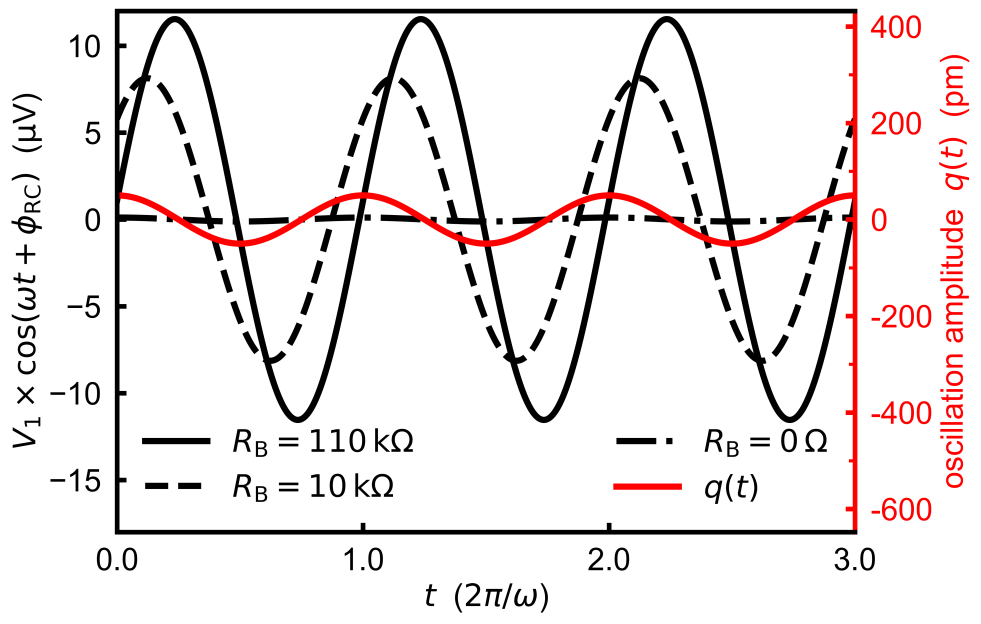}
	\caption{The AC component of $V_\text{s}(t)$ by Eq.~\eqref{eq_V_s_final} for $\langle I_\text{t}\rangle = - 1.0\,\text{nA}$ and different values of $R_\text{B}$. The oscillation amplitude of the tip is plotted for comparison. Parameters used here are $A=50\,\text{pm}$, $\kappa = 1\times10^{10}\,\text{m}^{-1}$, $C=761\,\text{pF}$, $\omega / (2\pi) = 20.4 \, \text{kHz}$. For a positive tunneling current, $V_1$ gains an additional minus sign which corresponds to mirroring at the x-axis. If the sensor bends by $180\,\text{pm}/V$ and $Q=200000$, an amplitude of $V_1=1\,\mu\text{V}$ at $\phi_\text{RC}=\mp90^\circ$ will lead to a change of the excitation signal of $\pm72\%$ at an amplitude setpoint of $A=50\,\text{pm}$.}
	\label{fig_Vs_plot}
\end{figure}

Both $\phi_{\text{RC}}$ and $V_1$ depend on the value of $R=R_\text{B}+R_\text{w}$, as illustrated in Fig.~\ref{fig_Vs_plot}. As $V_1 \propto R$, removing $R_\text{B}$ results not only in a significantly smaller phase shift (Eq.~\eqref{eq_phi_RC}) with respect to the oscillation $q(t)$ of the sensor, but also an equivalently reduced amplitude. Furthermore, $V_1$ depends linearly on the tunneling current as $I_{1\omega} \propto - \langle I_{\text{t}} \rangle$. \\
For a negative tunneling current $\langle I_{\text{t}} \rangle$, $V_{\text{s}}(t)$ lags behind the tip oscillation (see Fig.~\ref{fig_Vs_plot}) and the phase shift is $\phi_\text{RC}$. \\
For a positive tunneling current $\langle I_\text{t} \rangle$, the sign of the bias voltage $V_{\text{B}}$ is reversed; $V_{\text{s}}(t)$ gains an additional minus sign and is therefore ahead of the tip oscillation. The additional minus sign can also be interpreted as an additional phase shift of $180^\circ$: the phase shift of $V_{\text{s}}(t)$ with respect to the oscillation amplitude $q(t)=A\times\cos(\omega t)$ is then given by $\phi_{\text{RC}} + 180^\circ$. \\
With $\phi_{\text{RC}} = -85^\circ$, as estimated before in the case of $R=110.1\,\text{k}\Omega$, we see that for both positive and negative sign of $\langle I_\text{t}\rangle$, the phase shift is close to $+90^\circ$ or $-90^\circ$, respectively; $V_\text{s}(t)$ can thus strongly couple to the sensor oscillation as discussed in the next section.

\subsection{Harmonic oscillator model for the tip's motion \label{sec_HO_model}}

To study the influence of a piezoelectric force on the qPlus sensor motion, $q(t) = A\times\cos(\omega t)$, we consider a one-dimensional harmonic oscillator model. The equation of motion for the sensor tip with an effective mass $m^*$ and stiffness $k$ is \\
\begin{equation}
	m^*\,\ddot{q}(t)+\dfrac{m^* \omega_0}{Q}\,\dot{q}(t)+k \, q(t) = F_{\text{ts}}(t)+F_{\text{drive}}(t)+F_{\text{piezo}}(t), \label{eq_harmonic_osc}
\end{equation}
where $Q$ is the quality factor and $\omega_0 = 2\pi\times f_0 = \sqrt{k/m^*}$ is the angular resonance frequency of the free cantilever. External forces acting on the cantilever are written on the right-hand side of Eq.~\eqref{eq_harmonic_osc}: 1) $F_{\text{ts}}(t)$, the tip-sample force, 2) $F_{\text{drive}}(t) = F_{\text{drive}}^0 \times\cos(\omega t + \phi_{\text{drive}})$, the force of mechanical excitation of the sensor in FM-AFM mode, and 3) $F_{\text{piezo}}(t) = F_{\text{piezo}}^0 \times\cos(\omega t +\phi_{\text{piezo}})$, the piezoelectric force resulting from the modulation of the sensor voltage $V_{\text{s}}(t)$ which can interfere with the tip's motion and thereby produce an apparent dissipation. \\
The phases $\phi_{\text{drive}}$ and $\phi_{\text{piezo}}$ are defined in reference to the sensor oscillation $q(t)$. Since a voltage applied to the STM electrode results in an instantaneous force that deflects the sensor by the piezoelectric effect, we set $F_{\text{piezo}}(t)\propto V_1\times \cos(\omega t + \phi_\text{RC})$, and hence, $\phi_{\text{piezo}} = \phi_{\text{RC}}$. \\

All parameters on the left-hand side of Eq.~\eqref{eq_harmonic_osc} are well defined (see section~\ref{sec_exp_setup}). Both $F_{\text{drive}}(t)$ and $F_{\text{piezo}}(t)$ are assumed to be harmonic, more specifically, sinusoidal with the same angular frequency $\omega$ as the oscillation $q(t)$. For small deflections from the resting position $z_0$ or, equivalently, for small amplitudes of $q(t)$, one can assume a parabolic tip-sample interaction potential $V_\text{ts}$; then, the force gradient $\big(-\frac{\text{d}F_{\text{ts}}}{\text{d}z}\big)\big|_{z_0}=\big(\frac{\text{d}^2V_{\text{ts}}}{\text{d}z^2}\big)\big|_{z_0}$ is constant over one oscillation cycle and $F_{\text{ts}}(t)$ is a linear function of tip-sample distance $q(t)$ and thus also an harmonic function of time $t$:
\begin{equation}
	F_{\text{ts}}(t)=-k_{\text{ts}}(z_0)\times q(t) + const., \label{eq_F_ts}
\end{equation}
where $const.=0$, because a zero force would result in a zero deflection $q(t)$. Bringing $F_{\text{ts}}(t)$ to the left-hand side of Eq.~\eqref{eq_harmonic_osc}, dividing the whole equation by $m^*$ and using
\begin{equation}
	\omega^2 = \dfrac{k+k_{\text{ts}}}{m^*}, \label{eq_om^2}
\end{equation}
one ends up with the differential equation
\begin{eqnarray}
	\ddot{q}(t)+\dfrac{\omega_0}{Q}\,\dot{q}(t)+\omega^2\, q(t) &=& \dfrac{F_{\text{drive}}^0}{m^*}\times\cos(\omega t + \phi_{\text{drive}})\nonumber\\ 
	& &+\dfrac{F_{\text{piezo}}^0}{m^*}\times\cos(\omega t +\phi_{\text{RC}}). \nonumber \\
	\label{eq_HO_diffeq}
\end{eqnarray}
Inserting $q(t)=A\times\cos(\omega t)$, carrying out the derivatives and using a trigonometric identity for the cosine functions on the right-hand side, we have
\begin{eqnarray}
	&&\dfrac{-\omega \omega_0}{Q}\,A\times\sin(\omega t)= \nonumber\\
	&&\quad\bigg( \dfrac{F_{\text{drive}}^0}{m^*}\,\cos(\phi_{\text{drive}})+\dfrac{F_{\text{piezo}}^0}{m^*}\,\cos(\phi_{\text{RC}}) \bigg)\times\cos(\omega t) \nonumber\\
	&&\quad - \bigg( \dfrac{F_{\text{drive}}^0}{m^*}\,\sin(\phi_{\text{drive}})+\dfrac{F_{\text{piezo}}^0}{m^*}\,\sin(\phi_{\text{RC}}) \bigg)\times\sin(\omega t). \label{eq_0} \nonumber\\
\end{eqnarray}	
As sine and cosine are linearly independent, we can write
\begin{equation}
	\dfrac{\omega\omega_0}{Q}\, A=\dfrac{F_{\text{drive}}^0}{m^*}\,\sin(\phi_{\text{drive}})+\dfrac{F_{\text{piezo}}^0}{m^*}\,\sin(\phi_{\text{RC}}). \label{eq_1}
\end{equation}
At fixed $z_0$ and constant amplitude $A$, the left-hand side of Eq.~\eqref{eq_1} is constant. The two terms on the right-hand side are determined by the amplitudes of the two forces, $F_{\text{drive}}(t)$ and $F_{\text{piezo}}(t)$, that characterize the excitation of the sensor beam. To keep the amplitude $A$ and thus the left-hand side constant, these two terms have to cancel each other out, up to a constant offset given by the setpoint amplitude.\\

The parasitic piezoelectric excitation described by the force $F_{\text{piezo}}(t)$ depends on the presence of a tunneling current. In the limit of the free, mechanically driven cantilever with no tip-sample interaction forces ($\omega=\omega_0$ and $\phi_{\text{drive}}=90^\circ$) and zero tunneling current ($F_{\text{piezo}}(t)=0\,\text{N}$), Eq.~\eqref{eq_1} simplifies to
\begin{equation}
	\dfrac{F_{\text{drive}}^0}{m^*} =\dfrac{\omega_0^2}{Q}\,A|_\text{free},
\end{equation}
where $A|_\text{free}$ denotes the amplitude in the case of a free, driven oscillation, which is directly proportional to the amplitude excitation signal $X'_{\text{drive}}$:
\begin{equation}
	F_{\text{drive}}^0 =\dfrac{k}{Q}\,\alpha \, X'_{\text{drive}}, \label{eq_f_drive_final}
\end{equation}
where the proportionality factor $\alpha$ is obtained from experimental data (see section~\ref{sec_alphabeta}). \\

The piezoelectric force $F_{\text{piezo}}(t)$, on the other hand, is directly linked to the modulation of $V_{\text{s}}(t)$. We propose an elastic force with
\begin{equation}
	F_{\text{piezo}}(t) =\big(k+k_{\text{ts}}(z_0)\big)\times q_\text{pe}(t),
\end{equation}
where $q_\text{pe}(t)$ is the deflection of the cantilever due to the inverse piezoelectric effect caused by the AC voltage $V_1\times \cos(\omega t + \phi_\text{RC})$ applied to the STM electrode. For small deflections, the deformation of a piezoelectric crystal is directly proportional to the applied voltage:
\begin{equation}
	q_\text{pe}(t) = \beta \, V_1 \times \cos(\omega t + \phi_{\text{RC}})
\end{equation}
with the proportionality factor $\beta$, which can also be determined experimentally (see section~\ref{sec_alphabeta}).
As a result for $F_{\text{piezo}}^0$, one finds 
\begin{equation}
	F_{\text{piezo}}^0 = -k\,\dfrac{\omega^2}{\omega_0^2}\,\beta\,\dfrac{2\,R\,\langle I_{\text{t}}\rangle}{\sqrt{1+(\omega R C)^2}}\, \dfrac{\mathfrak{I}_1(2\kappa A)}{\mathfrak{I}_0(2\kappa A)}, \label{eq_f_piezo_final}
\end{equation}
where Eq.~\eqref{eq_om^2} and Eq.~\eqref{eq_I_t_Fourier_coeff} have been used. \\

To verify if our model agrees with experimental data, we combine the results found so far: from Eq.~\eqref{eq_1}, it follows that
\begin{eqnarray}
		A&=&\dfrac{Q}{k}\,\dfrac{\omega_0}{\omega}\,\big(F_{\text{drive}}^0 \,\sin(\phi_{\text{drive}})+F_{\text{piezo}}^0 \, \sin(\phi_{\text{RC}})\big)\nonumber\\
		&=&\dfrac{\omega_{0}}{\omega}\,\alpha\, X'_{\text{drive}}\,\sin(\phi_{\text{drive}})
		+Q\,\dfrac{\omega}{\omega_{0}}\,\beta\,\dfrac{2R\langle I_{\text{t}}\rangle}{\sqrt{1+(\omega R C)^2}}\nonumber\\
		& &\times\dfrac{\mathfrak{I}_1(2\kappa A)}{\mathfrak{I}_{0}(2\kappa A)}\sin\big(\arctan(\omega R C)\big), \label{eq_A_HO}
\end{eqnarray}
where $\phi_{\text{RC}}$, $F_{\text{drive}}^0$ and $F_{\text{piezo}}^0$ have been inserted from Eq.~\eqref{eq_phi_RC}, Eq.~\eqref{eq_f_drive_final} and Eq.~\eqref{eq_f_piezo_final}, respectively. \\ 
Equivalently, one can solve Eq.~\eqref{eq_1} for $X'_{\text{drive}}$, which results in
\begin{eqnarray}
	X'_{\text{drive}}&=&\dfrac{\omega}{\omega_{0}}\,\dfrac{A}{\alpha \,\sin(\phi_{\text{drive}})} -Q\,\dfrac{\omega^2}{\omega_{0}^2}\,\dfrac{\beta}{\alpha}\, \dfrac{2R\langle I_{\text{t}}\rangle}{\sqrt{1+(\omega R C)^2}}\nonumber\\
	& &\times\dfrac{\mathfrak{I}_1(2\kappa A)}{\mathfrak{I}_{0}(2\kappa A)}\, \dfrac{\sin\big(\arctan(\omega R C)\big)}{\sin(\phi_{\text{drive}})}. \label{eq_X_HO}
\end{eqnarray}
The two latter equations show the interplay between amplitude $A$, excitation $X'_{\text{drive}}$ and tunneling current $\langle I_{\text{t}} \rangle$. The amplitude $A$ (Eq.~\eqref{eq_A_HO}) is constant if the changes in $F_\text{drive}^0$ and $F_\text{piezo}^0$ compensate each other. Nevertheless, since $X'_\text{drive}$ (Eq.~\eqref{eq_X_HO}) is linear in $\langle I_\mathrm{t} \rangle$ (see experimental data in section~\ref{sec_piezophenom}), both terms of the sum on the right-hand side of Eq.~\eqref{eq_A_HO} depend linearly on $\langle I_{\text{t}} \rangle$ and thus exponentially on $z$. At large enough tunneling currents, where the PI-controller cannot adjust the excitation $X'_{\text{drive}}$ fast enough during a $z$-sweep measurement, the amplitude $A$ will change accordingly.

\section{Testing the harmonic model \label{sec_test}}

In this section, the results of the theoretical model presented previously are tested by comparison with measured data from experiment. \\ 
The harmonic model (section~\ref{sec_harmonic_model}) is verified by comparing the results from Eq.~\eqref{eq_X_HO} with experimental values of the excitation $X'_{\text{drive}}$ during $z$-spectroscopy. More specifically, we compare values for $\Delta E_\text{ts}^\text{apparent} / \langle I_\text{t} \rangle$, which are obtained from a linear fit of $\Delta E_\text{ts}^\text{apparent}$ versus $\langle I_\text{t} \rangle$: $\Delta E_\text{ts}^\text{apparent}$ can be calculated (Eq.~\eqref{eq_dE_apparent}) using either the measured excitation signal or theoretical values from Eq.~\eqref{eq_X_HO}; the data (phase $\phi_\text{drive}$, angular frequency $\omega$ and amplitude $A$) needed for calculation of the theoretical values as a function of the tunneling current $\langle I_\text{t}\rangle$ stems from corresponding $z$-spectroscopy measurements. The decay constant $\kappa$ can easily be determined: fitting $\langle I_{\text{t}} \rangle$ linearly versus $z$ on a logarithmic scale, results in a straight line with slope $(-2\kappa)$. \\
The remaining free parameters of Eq.~\eqref{eq_X_HO} are $\alpha$ and $\beta$, which are discussed in the following.

\subsection{Fixing the parameters for the harmonic model \label{sec_alphabeta}}

\begin{figure}
	\centering
	\includegraphics{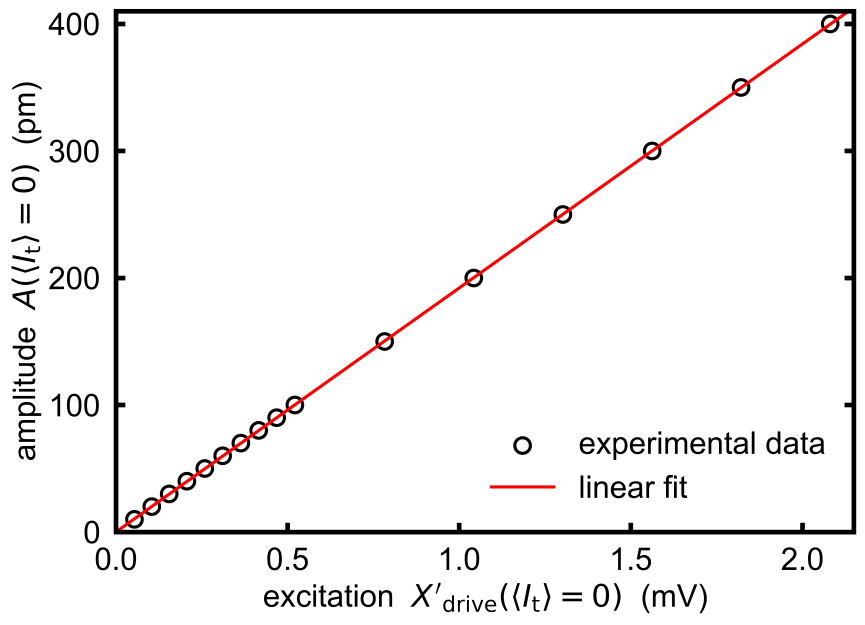}
	\caption{Determination of $\alpha$ from data of $z$-spectroscopy measurement. Values for both amplitude $A$ and excitation $X'_\text{drive}$ for $\langle I_\text{t} \rangle = 0 \,\text{nA}$ (which corresponds, to a good approximation, to the case of the free oscillation) are obtained from linear fit of $A$ and $X'_\text{drive}$, respectively, as a function of $\langle I_\text{t}\rangle$.
		\label{fig_ape}}
\end{figure}

The amplitude of the driving force $F_{\text{drive}}^0$ depends on the ratio $\alpha$ of the amplitude $A|_\text{free}$ for the case of a free oscillation to the excitation $X'_{\text{drive}}$ (see Eq.~\eqref{eq_f_drive_final}). The amplitude response to the mechanical excitation using the piezo-tube simultaneously depends on the overall expansion of the piezo-tube and $\alpha$ is therefore different in the case of a fully withdrawn tip than in the regime, where our $z$-spectroscopy data was measured. As the excitation $X'_{\text{drive}}$ changes according to the apparent dissipation, while conservative forces like the tip-sample interaction only change the frequency shift $\Delta f$, leaving amplitude $A$ and excitation $X'_{\text{drive}}$ constant, the value of $\alpha$ can also be obtained from the ratio of the measured amplitude $A$ to the excitation $X'_{\text{drive}}$ in the case of a finite tip-sample force (finite frequency shift $\Delta f$), but vanishing tunneling current $\langle I_{\text{t}}\rangle$. Plotting amplitude $A$ and excitation $X'_\text{drive}$, respectively, as a function of $\langle I_\text{t} \rangle$ (from the $z$-spectroscopy data), values of $A$ and $X'_\text{drive}$ for $\langle I_{\text{t}}\rangle = 0\,\text{nA}$ can be obtained as the y-offset of a linear fit in both cases; the value of $\alpha$ is in turn given by the slope of a linear fit of these values (Fig.~\ref{fig_ape}):
\begin{equation}
	\alpha = 192.1 \,\text{nm}/\text{V}. \label{eq_alpha_value}
\end{equation}

\begin{figure}
	\centering
	\includegraphics{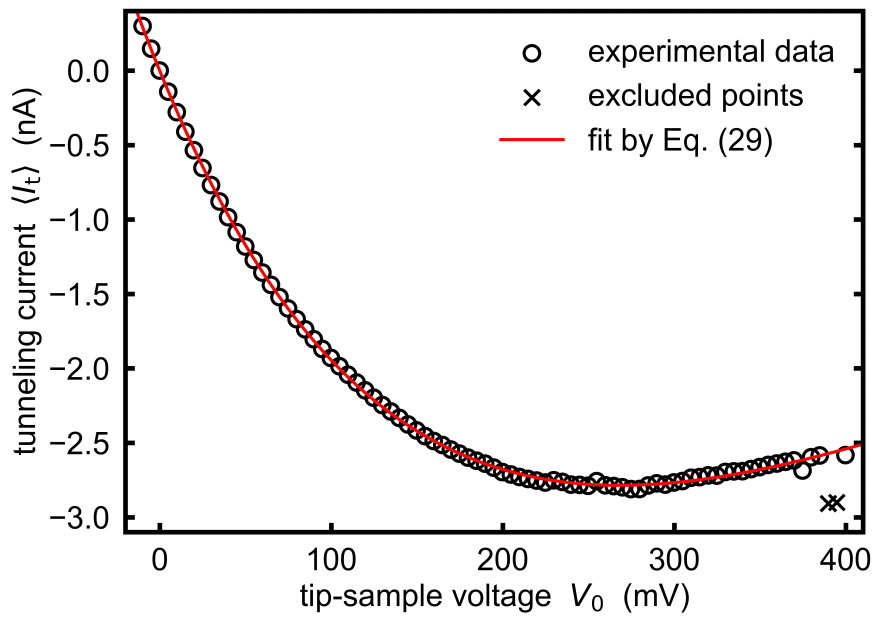}
	\caption[Tunneling current $\langle I_{\text{t}}\rangle$ versus voltage $V_0$ between tip and sample.]{Tunneling current $\langle I_{\text{t}}\rangle$ versus voltage $V_0$ between tip and sample. From a fit of experimental data by Eq.~\eqref{eq_IV_bias_sweep} the value of $\beta$ can be obtained. Two data points have been excluded from the fit, which strongly differ from the other data points because of an instability of the tip at high voltages $V_0$ likely causing a change in the atomic configuration of the tip and thus an abrupt change in the tunneling current $\langle I_{\text{t}} \rangle$.
		\label{fig_bias_sweep}}
\end{figure}

The amplitude of the piezoelectric force $F_{\text{piezo}}^0$, on the other hand, depends on the value of $\beta$, the sensor deflection per potential at the STM electrode. It can be obtained from a bias voltage sweep as explained in the following.\\
The tunneling current as a function of the tip-sample distance $z$ is given in Eq.~\eqref{eq_tunneling_current}, where $I_0 = G_0\,V$; $G_0$ is the conductance quantum and $V$ is the potential difference between tip and sample. Since in our simplified model the sample lies at $V_\text{g}$, which is at ground, $V_{\text{s}}(t)$ equals the voltage between the sensor tip and the sample; in section~\ref{sec_Vs_calculation} (with appendix~\ref{app_Vs_harm}) it has been shown that the time-average of this potential is given by $V_0 = V_{\text{B}} + R \, \langle I_{\text{t}} \rangle$. \\
By the inverse piezoelectric effect, a tip-sample voltage $V_0$ leads to a change in tip-sample distance: $z \rightarrow z+\beta \, V_0$. For a given value of $z$, the tunneling current as a function of the voltage between tip and sample is therefore
\begin{equation}
	I_{\text{t}}(V_0)=G_0\, V_0\times\exp\big(-2\kappa (z + \beta \, V_0)\big). \label{eq_IV_bias_sweep}
\end{equation}

From a bias voltage sweep in constant height mode experimental values for $I_\text{t}(V_0)$ are obtained, which can be fitted by Eq.~\eqref{eq_IV_bias_sweep}. The decay constant $\kappa$ is determined from a separate $z$-spectroscopy measurement: $\kappa = 1.033 \times 10^{10} \, \text{m}^{-1}$.
The fit in Fig.~\ref{fig_bias_sweep} results in
\begin{equation}
	\beta = 180.5 \, \text{pm}/\text{V}.
\end{equation}
 
\subsection{Results \label{sec_results}}

Experimental values for $\Delta E_\text{ts}^\text{apparent} / \langle I_\text{t} \rangle$ can be obtained from linear fit of $\Delta E_\text{ts}^\text{apparent}$ as a function of $\langle I_\text{t}\rangle$. A theoretical result from our model can be obtained by using $X'_\text{drive}$ as calculated by Eq.~\eqref{eq_X_HO}, instead of experimental data to obtain $\Delta E_\text{ts}^\text{apparent}$ (Eq.~\eqref{eq_dE_apparent}), as discussed at the beginning of section~\ref{sec_test}; after $\alpha$ and $\beta$ have been determined, no free parameters are left. \\

\begin{figure}
	\centering
	\includegraphics{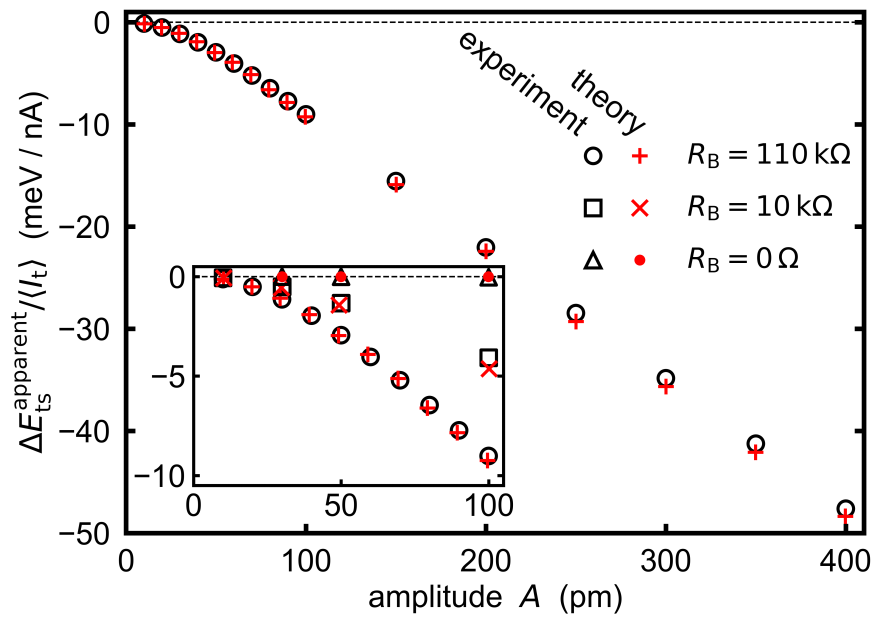}
	\caption{$\Delta E_\text{ts}^\text{apparent} / \langle I_\text{t}\rangle$ as a function of amplitude $A$. $\Delta E_\text{ts}^\text{apparent}$ is obtained from $X'_\text{drive}$ (experimental values and theoretical calculation by Eq.~\eqref{eq_X_HO}) using Eq.~\eqref{eq_dE_apparent}. The theoretical results start to deviate from experiment at $\kappa A \approx 1$ (i.e., $A \approx 100\,\text{pm}$), where the harmonic approximation is no longer valid.
		\label{fig_results}}
\end{figure}

Fig.~\ref{fig_results} shows a comparison between experiment and theory for multiple oscillation amplitudes $A$ between 10\,pm and 400\,pm for $R_\text{B}=110\,\text{k}\Omega$. The inset graph additionally shows a comparison for smaller values of $R_\text{B}$, which were also adopted in measurement. As the results, for given $A$ and $R_\text{B}$, are independent of the ramping direction and the magnitude of the bias $|V_\text{B}|$, it is valid to consider the average for both ramping directions and all bias voltages. From the experimental data one finds that $\Delta E_\text{ts}^\text{apparent}/\langle I_{\text{t}} \rangle$ is reduced to around $45 \,\%$ for $R_\text{B}=10\,\text{k}\Omega$ and to less than $1\,\%$ for $R_\text{B}=0\,\Omega$ (no external resistor) compared to the value observed with $R_\text{B}=110\,\text{k}\Omega$, respectively (for exemplary values see appendix~\ref{app_piezo_creep} with Table~\ref{tab_diss_diffRB}). \\
Since $\Delta E_\text{ts}^\text{apparent}$ and $\langle I_\text{t} \rangle$ have opposite sign (as discussed in section~\ref{sec_piezophenom}), $\Delta E_\text{ts}^\text{apparent} / \langle I_\text{t} \rangle$ is always negative. For amplitudes $A$ below $100\,\text{pm}$, the behavior of $\Delta E_\text{ts}^\text{apparent} / \langle I_\text{t} \rangle$ is dominated by the quadratic term $E\propto A^2$ in Eq.~\ref{eq_dE_apparent}, while for larger amplitudes the factor $(X'_\text{drive}/X_\text{drive}-1)$ increases more significantly, such that the overall behavior becomes almost linear.\\
In general, very good agreement is found for the theoretical values compared to experiment. For $A=10\,\text{pm}$ or $R_\text{B}=0\,\Omega$, $\Delta E_\text{ts}^\text{apparent} / \langle I_\text{t} \rangle$ is close to zero, such that the theoretical results lie within the noise level of the experiment. The largest relative error for points with $A>10\,\text{pm}$ and $R_\text{B}>0\,\Omega$ is found for $A=100\,\text{pm}$, $R_\text{B}=10\,\text{k}\Omega$ and is 14\%; around $\kappa A \approx 1$ ($A\approx 100\,\text{pm}$), the harmonic approximation of our model is no longer valid (see also appendix~\ref{app_Vs_solution}), such that the values deviate from experiment. Nevertheless, for $R_\text{B} = 110\,\text{k}\Omega$, the largest relative error for $A\geq 100\,\text{pm}$ is only 3\%.

\subsection{Possible solutions to minimize the cross coupling\label{sec_sol}}

Possible solutions to the presented problem of cross coupling are very briefly discussed in the following.\\

Since the effect of a cross coupling between the sensor excitation and the tunneling current inherently relies on the electrode design of the qPlus sensor in our microscope, an alternative sensor design is conceivable, e.g., where the tip is electrically isolated from the quartz sensor and is contacted via a gold wire as suggested in Ref.~\onlinecite{Majzik2012}. The oscillating potential at the tip then has no direct contact to the quartz cantilever, severely decreasing the piezoelectric force acting on the sensor. Nonetheless, as discussed in Ref.~\onlinecite{Nony2016}, the oscillating tip and the modulation of the potential at the tip acts as an electric dipole, emitting electromagnetic waves that again can couple to the excitation of the qPlus sensor via the inverse piezoelectric effect. Nevertheless, the extra wire can lead to instability of the $Q$ factor and multiple resonances. \\
Another possible solution would be to modulate the bias voltage signal at the sensor using the same signal phase-shifted by $180^\circ$ in order to completely cancel the oscillation of the potential at the STM electrode (negative feedback): inductors, like coils, could be used in order to shift the phase intentionally. \\
Removing the resistance $R$ is probably the easiest solution, in theory; in practice, however, a finite resistance of the wiring will always remain. As we have seen in section~\ref{sec_results}, the dissipated energy decreases drastically when removing $R_\text{B}$: the apparent dissipation is negligibly small compared to the case with $R_\text{B}=110\,\text{k}\Omega$. The big disadvantage of this approach is that an alternative kind of noise filter for the bias voltage signal is required since removing $R_\text{B}$ also eliminates the RC low-pass filter element in the circuit. One solution to this is adding a buffer amplifier after the RC filter such that the modulation of the tunneling current cannot cause a significant oscillation of the bias voltage. \\
A different solution is provided by Schwenk et al. in Ref.~\onlinecite{Schwenk2021}, who provide means to lift the potential of the AFM amplifier to the bias voltage, so all sensor electrodes refer to that bias voltage.

\section{Summary}

The origin of cross coupling between tunneling current and excitation in a home-built, combined scanning tunneling and atomic force microscope setup has been identified. In dynamic AFM operation modes such as frequency modulation AFM, where the tip oscillates, the tunneling current flowing through the tip at the sensor will also oscillate as a function of time. If the impedance of the bias voltage supply is significant, such as when using a low-pass filter without a buffer, the oscillating current will cause an oscillation in the bias voltage, leading to a cross coupling between excitation and tunneling current, which has been observed and discussed previously on several different occasions. In this study, the origin of the cross coupling was investigated using a new approach for a quantitative description. This approach is based on the inverse piezoelectric effect and accounts for the harmonic modulation of the potential at the tip electrode. In our setup, this voltage oscillation is caused by an RC low-pass filter in the experimental setup consisting of a resistance $R$ in the cable that connects the bias output to the sensor, and the capacitance $C$ between the STM-electrode of our qPlus sensor and ground; this low-pass filter shifts the oscillating potential on the STM-electrode to be out of phase with the sensor oscillation. The measured apparent dissipation signal is linked to a piezoelectric force which is proportional to the oscillating component of that potential. We have shown this by implementing the harmonic oscillator model to describe the motion of the sensor cantilever. Within an error margin of 14\%, the results agree well with the experimental data from $z$-spectroscopy measurements.

\begin{acknowledgments}
	We thank Ferdinand Huber and Julian Berwanger for providing initial current versus damping data and Jay Weymouth for discussions. MS further thanks Raphael Lehner, who helped him with some calculations. 
	This publication is strongly shortened version of the bachelor thesis of MS, that was guided by FS and MW and supervised by FJG.
	We also thank the Deutsche Forschungsgemeinschaft for funding under CRCs 689 and 1277. 
\end{acknowledgments}

\section*{Author declarations}

\subsection*{Conflict of Interest}
FJG holds patents about the force sensor
that was used in the experiments. The remaining authors have no
conflicts to disclose.

\section*{Data availability}
The data that support the findings of
this study are available from the
corresponding author upon reasonable
request.

\section*{Appendix}

\appendix

\section{Supplementary experimental data \label{app_addplots}}

\subsection{Testing the virtual ground of the STM pre-amplifier \label{app_Vg}}

\begin{figure}
	\centering
	\includegraphics{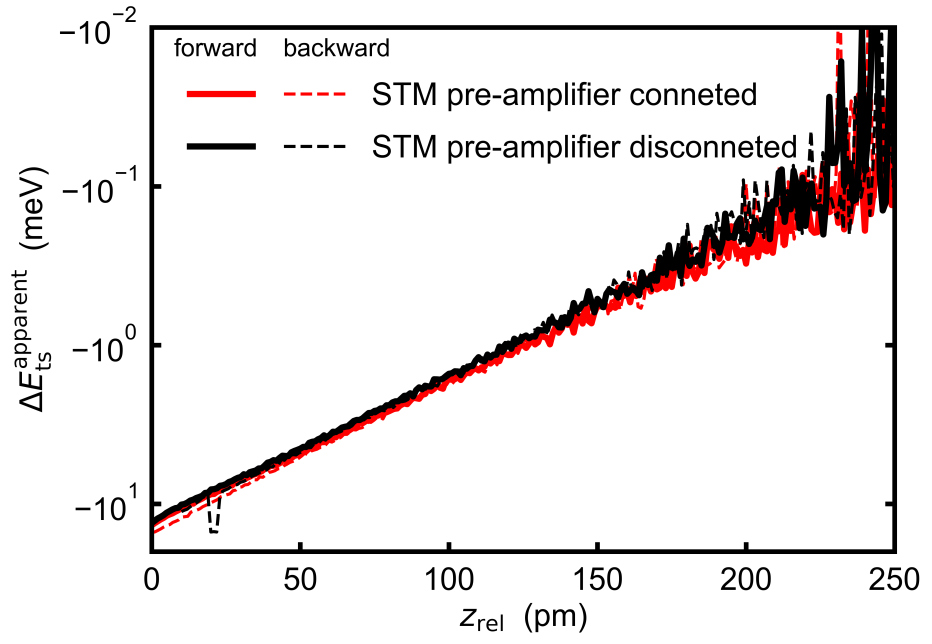}
	\caption{Comparison of $\Delta E_\text{ts}^\text{apparent}$ for measurements with the STM pre-amplifier connected (as depicted in Fig.~\ref{fig_STM_electronics}) as well as the amplifier disconnected and the sample grounded. The relative tip height $z_\text{rel}$ was chosen as x-axis since the tunneling current cannot be measured with the pre-amplifier disconnected. The discrepancy in the backward (dashed) lines can be attributed to piezo creep and a separate drift compensation in both cases as well as an instable tip (which is evident from the step in the black dashed line at around $z_\text{rel}=10\,\text{pm}$). Since the red and black curves agree reasonably well, setting $V_\text{g}$ to ground is valid.}
	\label{fig_STMpreamp_check}
\end{figure}

In section~\ref{sec_exp_setup} it is stated that the apparent dissipation $\Delta E_\text{ts}^\text{apparent}$ with the STM pre-amplifier disconnected is equal to $\Delta E_\text{ts}^\text{apparent}$ in the case with the amplifier connected, such that $V_\text{g}$ can assumed to be at ground. For the two cases (with and without the STM pre-amplifier), Fig.~\ref{fig_STMpreamp_check} shows a comparison of the apparent dissipation $\Delta E_\text{ts}^\text{apparent}$ as a function of the relative tip height $z_\text{rel}$ for forward and backward direction, respectively. With the STM pre-amplifier disconnected the tip is approached in AFM feedback using the same frequency shift setpoint at $z_\text{rel}=0\,\text{pm}$ as for the measurement with the STM pre-amplifier connected. The forward ramps in both measurements match well. Because of piezo creep during the measurement, however, a small discrepancy between forward and backward ramps arises, which differs for both cases due to separate drift-compensation (see also appendix~\ref{app_piezo_creep}). This effect was reduced by adjusting the offset and range of the $z_\text{rel}$ values in accordance with the df-signal of the corresponding forward ramp. Tip instability caused a step at around $z_\text{rel}=10\,\text{pm}$ in the backward measurements with the STM pre-amplifier disconnected. Overall, the apparent dissipation $\Delta E_\text{ts}^\text{apparent}$ in the two cases matches well for both ramping directions; assuming $V_\text{g}=0\,\text{V}$ is reasonable.

\subsection{Different $R_\text{B}$, ramping directions, and piezo creep \label{app_piezo_creep}}

\begin{figure}
	\centering
	\includegraphics{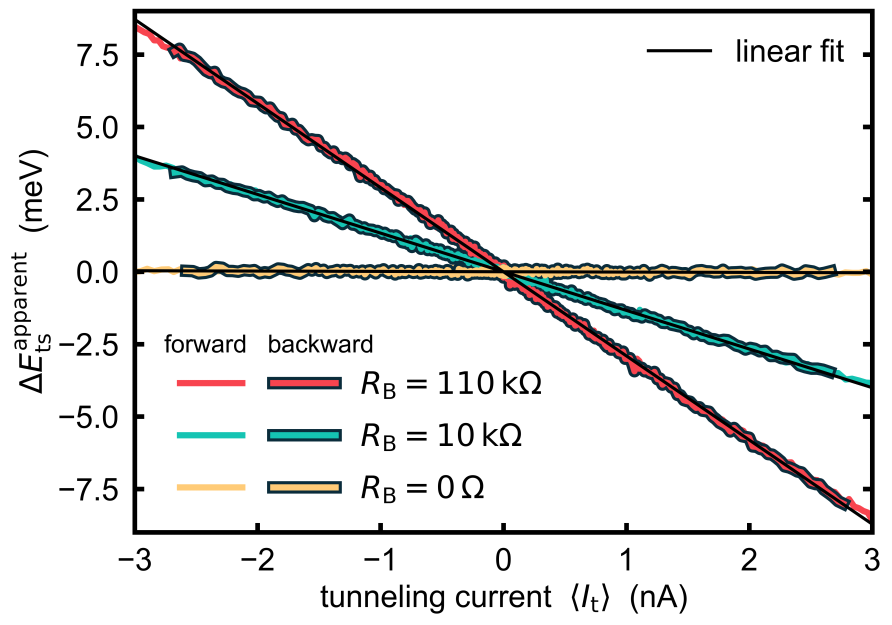}
	\caption{$\Delta E_\text{ts}^\text{apparent}$ for different values of $R_\text{B}$, for forward and backward ramping direction of the $z$-spectroscopy, respectively, for $A=50\,\text{pm}$ and $V_\text{B}=\pm1.0\,\text{mV}$. The slopes of the curves as obtained from linear fit correspond to the values in Table~\ref{tab_diss_diffRB}. The graphs also make it clear that $\Delta E_\text{ts}^\text{apparent} / \langle I_\text{t} \rangle$ is in general independent of the ramping direction during measurement.}
	\label{fig_dE_diffRB}
\end{figure}

In Fig.~\ref{fig_dE_diffRB}, $\Delta E_\text{ts}^\text{apparent}$ is plotted as a function of tunneling current $\langle I_\text{t} \rangle$ for forward and backward ramping direction for an amplitude of $A=50\,\text{pm}$ and three different values of $R_\text{B}$. A linear fit of the curves for each of the three values of $R_\text{B}$, respectively, gives the values for $\Delta E_\text{ts}^\text{apparent} / \langle I_\text{t} \rangle$ listed in Table~\ref{tab_diss_diffRB}. \\ 
From the graphs (Fig.~\ref{fig_dE_diffRB}) is also evident that $\Delta E_\text{ts}^\text{apparent}$ is independent of the ramping direction during measurement as stated in section~\ref{sec_piezophenom}. This is valid as long as the ramping speed is chosen slow enough for the PI-controller to be able to effectively control the excitation to keep the oscillation amplitude constant as discussed at the end of section~\ref{sec_HO_model}. \\
The maximal tunneling currents in the backward ramps are about 10\% smaller compared to the forward ramps as during measurement a drift in the $z$-position of the tip, for example due to piezo-creep, can occur. When the tip is moved $500\,\text{pm}$ away from and again $500\,\text{pm}$ towards the sample during the $z$-spectroscopy measurement, the value of the tunneling current at the end will differ from the value at the beginning of the measurement by 10\% if the total distance $z$ has merely changed by 5\,pm. This effect was minimized by performing a drift-compensation before each measurement, which helps, e.g., for thermal drift, but not for the piezo-creep originating from the $z$-position movement during the $z$-spectroscopy experiments. 

\begin{table}
	\caption{\label{tab_diss_diffRB} Values for the apparent dissipated energy per oscillation cycle and tunneling current $\Delta E_{\text{ts}}^{\text{apparent}}\, \big/ \, \langle I_{\text{t}}\rangle$ obtained from measurements with three different resistors $R_\text{B}$ in the bias line for an amplitude $A=50\,\text{pm}$ (see fits in Fig.~\ref{fig_dE_diffRB}).}
	\begin{ruledtabular}
		\begin{tabular}{cc}
			resistance $R_\text{B}$  (k$\Omega$) & $\Delta E_{\text{ts}}^{\text{apparent}} / \langle I_{\text{t}}\rangle$  (meV/nA) \\ 
			
			\hline
			
			$110$ & $-2.95$ \\ 
			
			$10$ & $-1.33$ \\ 
			
			$0$ & $-0.01$ \\
		\end{tabular}
	\end{ruledtabular}		
\end{table}

\section{Calculation of sensor potential and validity of the harmonic model \label{app_Vs_solution}}

In this section the sensor potential $V_\text{s}(t)$ (Eq.~\eqref{eq_V_s_final}) as used in the harmonic model of section~\ref{sec_harmonic_model} is calculated. \\
Furthermore, we show that the assumption of a harmonic tunneling current $I_\text{t}(t)$ is justified by explicitly performing the calculation with the tunneling current given by Eq.~\eqref{eq_tunneling_current_osc}.\\

The defining differential equation for the sensor potential $V_{\text{s}}(t)$, Eq.~\eqref{eq_V_s_diff_eq}, is an inhomogeneous linear differential equation of the form
\begin{equation}
	\dfrac{\text{d}V_{\text{s}}(t)}{\text{d}t}=a(t)\, V_{\text{s}}(t)+b(t), \label{eq_diff_eq}
\end{equation}
where $a$ and $b$ are continuous functions of time $t$ given by
\begin{equation}
	a(t)= -1/RC \quad \text{and} \quad b(t)= I_{\text{t}}(t)/C + V_{\text{B}}/RC.
\end{equation}
The unique solution to the differential equation Eq.~\eqref{eq_diff_eq} with initial condition $V_{\text{s}}(t_0)=c$ ($t_0, \, c \in \mathbb{R}$) is given by \cite{Forster.2017}
\begin{equation}
	V_{\text{s}}(t)=V_0(t)\bigg( c+ \displaystyle\int_{t_0}^{t} V_0(t')^{-1}\,b(t') \,\text{d}t' \bigg). \label{eq_variation}
\end{equation}
Here,
\begin{equation}
	V_0(t) = \exp\bigg( \displaystyle\int_{t_0}^{t} a(t')\,\text{d}t' \bigg)
\end{equation}
is the solution to the homogeneous differential equation, i.e., Eq.~\eqref{eq_diff_eq} with $b(t) = 0$. \\
If we choose the initial time as $t_0=0$, we have
\begin{equation}
	V_0(t)=\exp(-t / RC) \quad \text{and} \quad V_0(t)^{-1}=\exp(t/RC). 
\end{equation}
Putting everything together in Eq.~\eqref{eq_variation} gives
\begin{eqnarray}
	V_{\text{s}}(t) &=& \big( V_{\text{s}}(0)-V_{\text{B}} \big)\times \exp(-t/RC) + V_{\text{B}} \nonumber \\
	& & + \exp(-t/RC) \times \displaystyle\int_0^{t} \exp(t'/RC)\,I_{\text{t}}(t') / C \,\text{d}t'
	, \nonumber\\ \label{eq_V_s_inharm}
\end{eqnarray}
where $I_\text{t}(t')$ denotes the time-dependent tunneling current as discussed in section~\ref{sec_tunneling_current}.

\subsection{Harmonic tunneling current \label{app_Vs_harm}}

Before turning to the more rigorous calculation, the current is considered only up to first order in $\omega$,
\begin{equation}
	I_\text{t}(t) = \langle I_\text{t} \rangle + I_{1\omega} \times \cos(\omega t).
\end{equation}
Inserting this into Eq.~\eqref{eq_V_s_inharm} and using Euler's formula to solve the integral yields
\begin{eqnarray}
	V_{\text{s}}(t) &=& \bigg[ V_{\text{s}}(0)-\bigg(V_{\text{B}}+R\,\langle I_{\text{t}} \rangle + \dfrac{R\,I_{1\omega}}{1+(\omega R C)^2} \bigg) \bigg] \nonumber\\
	& & \times \exp(-t/RC) + V_{\text{B}} + R\,\langle I_\text{t} \rangle \times v_\text{s}(t),
	\label{eq_Vs_final_all}
\end{eqnarray}
where
\begin{eqnarray}
	v_\text{s}(t) &=& \dfrac{R\, I_{1\omega} \, \cos\big(\omega t - \arctan(\omega R C)\big)}{\sqrt{1+(\omega R C)^2}}
\end{eqnarray}
Leaving out the transient terms, that decay exponentially with time $t$, one finds the steady-state solution for $V_{\text{s}}(t)$ as stated in Eq.\eqref{eq_V_s_final}.

\subsection{Inharmonic tunneling current \label{app_Vs_inharm}}

\begin{figure}
	\centering
	\includegraphics{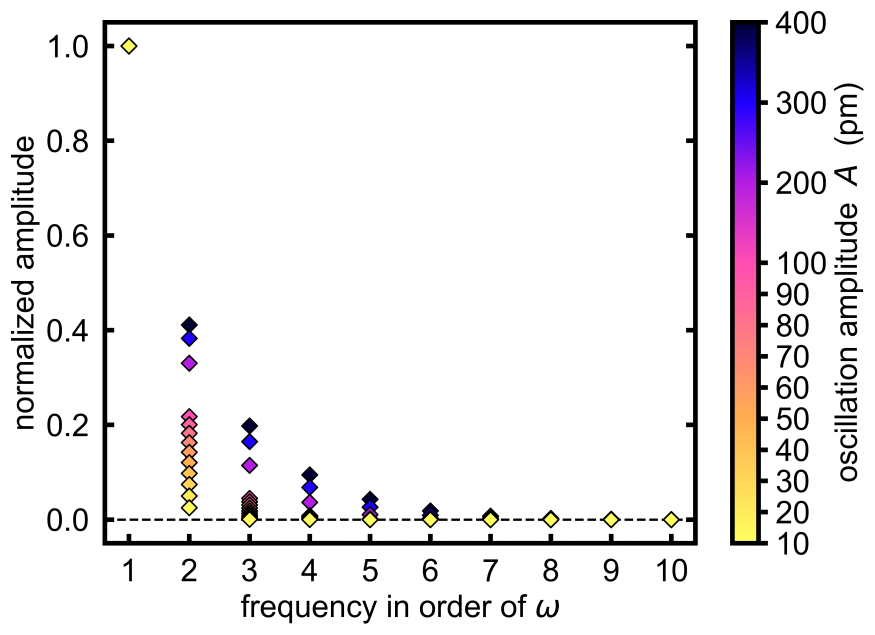}
	\caption{Amplitude of the first and higher order terms in $u_\text{s}(t)$ sorted by order in $\omega$ and normalized to the amplitude of the first order term, for different oscillation amplitudes $A$. Parameters used for the plot are $\kappa = 1\times 10^{10}\,\text{m}^{-1}$, $\omega/(2\pi) = 20.4\,\text{kHz}$, $R=110.1\,\text{k}\Omega$ and $C=761\,\text{pF}$. For $\kappa A \geq 1$ ($A\geq 100\,\text{pm}$) the second order term is larger than one fifth of the first order term; here, the harmonic approximation breaks down.}
	\label{fig_us_vs_FFT}
\end{figure}

To show that assuming a harmonic behavior of $V_{\text{s}}$ is valid, the sensor potential can also be calculated by inserting the actual tunneling current,
\begin{eqnarray}
	I_{\text{t}}(t) &\stackrel{\text{Eq.~\eqref{eq_tunneling_current_osc}}}{=}& I_{\text{z}_0}\times\exp\big(-2\kappa A\cos(\omega t)\big)\nonumber\\
	&\stackrel{\text{Eq.~\eqref{eq_I_t_Fourier_coeff}}}{=}& \dfrac{\langle I_{\text{t}} \rangle}{\mathfrak{I}_0(2\kappa A)}\times\exp\big(-2\kappa A\cos(\omega t)\big), \quad \label{eq_I_t(t)}
\end{eqnarray}
into Eq.~\eqref{eq_V_s_inharm}. Using again Euler's formula, the series expansion of the exponential, and 
\begin{equation}
	\mathfrak{I}_l(2x) = \displaystyle \sum_{n=0}^{\infty} \dfrac{x^{2n+l}}{n!(n+l)!}\qquad (x\geq0), 
	\label{eq_besselfct}
\end{equation}
the result can be rewritten: 
\begin{eqnarray}
	V_{\text{s}}(t) &=& \big( V_{\text{s}}(0)-V_{\text{B}}-R\,\langle I_{\text{t}} \rangle \big) \times \exp(-t/RC) \nonumber\\ 
	& &+V_{\text{B}}+R\,\langle I_{\text{t}} \rangle+u_{\text{s}}(t), \label{eq_V_s_inharm2}
\end{eqnarray}
where
\begin{eqnarray}
	u_{\text{s}}(t)&=& \dfrac{2 \, R \,\langle I_{\text{t}} \rangle}{\mathfrak{I}_0(2\kappa A)}\,\displaystyle\sum_{\substack{n=0\\m = 1\\n < m}}^{\infty} \dfrac{(- \kappa A)^{n+m}}{n! \, m! \times \left[1+\big(\omega RC\,(m-n)\big)^2\right]} \nonumber \\
	& &\times\Big\{ \cos\big(\omega\,t\,(m-n)\big) - \exp(-t/RC) \nonumber \\
	& & \qquad \qquad + \omega RC\,(m-n)\times\sin\big( \omega\,t\,(m-n) \big) \Big\}.  \nonumber \\ 
	\label{eq_u'_s(t)}	
\end{eqnarray}
Neglecting all terms with an exponentially decaying factor, one has
\begin{equation}
	V_{\text{s}}(t) = V_{\text{B}}+R\,\langle I_{\text{t}} \rangle+u_{\text{s}}(t),
\end{equation} 
and
\begin{eqnarray}
	u_{\text{s}}(t)&=& \dfrac{2\, R \,\langle I_{\text{t}} \rangle}{\mathfrak{I}_0(2\kappa A)}\,\displaystyle\sum_{\substack{n=0\\m = 1\\n < m}}^{\infty} \dfrac{(- \kappa A)^{n+m}}{n! \, m!} \nonumber \\
	& &\times \dfrac{\cos\left[\omega t\,(m-n) - \arctan\big(\omega RC\,(m-n)\big) \right]}{\sqrt{1+\big( \omega RC\,(m-n)\big)^2}}. \nonumber \\ 
	\label{eq_u_s(t)}
\end{eqnarray}
Terms of first order in $\omega$ correspond to terms of the sum where $m = n+1$, second order terms correspond to $m = n+2$, and so on; hence, with Eq.~\eqref{eq_besselfct} we find
\begin{equation}
	u_{\text{s}}(t) = v_{\text{s}}(t) + \displaystyle \sum_{l=2}^{\infty} \dfrac{R \, I_{l\omega} \, \cos\big(l\omega t - \arctan(l\omega RC)\big)}{\sqrt{1+(l\omega R C)^2}},
\end{equation}
where
\begin{equation}
	I_{l\omega} = (-1)^l\times2\,\langle I_{\text{t}} \rangle \, \dfrac{\mathfrak{I}_l(2\kappa A)}{\mathfrak{I}_0(2\kappa A)}, \qquad l=1,2,3,...
\end{equation}
are the Fourier components of the tunneling current of first and higher orders (see section~\ref{sec_tunneling_current}). \\
Second and higher order terms in $u_\text{s}(t)$ are small compared to $v_\text{s}(t)$ if $\kappa A < 1$ since then the sum in Eq.~\eqref{eq_u_s(t)} converges quickly. \\
Fig.~\ref{fig_us_vs_FFT} shows the amplitudes of the terms in $u_\text{s}(t)$ up to tenth order in $\omega$ for different oscillation amplitudes $A$ between 10\,pm and 400\,pm, normalized to the amplitude of the first order term $v_\text{s}(t)$. \\
For $A < 100\,\text{pm}$ already the fifth order term has a negligible amplitude, for $A = 400\,\text{pm}$ only the tenth and higher order terms are negligibly small. 
Adopting a harmonic oscillation of the tunneling current $I_{\text{t}}(t)$ for the calculation of the sensor potential $V_{\text{s}}(t)$ is therefore justified; the equation of motion of the harmonic oscillator, including the assumption of a harmonic (piezoelectric) dissipative force $F_{\text{piezo}}(t)$, is an appropriate way of describing the dynamic motion of the tip affected by parasitic dissipation if $\kappa A < 1$. 
For larger tip amplitudes $A$ the harmonic approximation breaks down as the second order term is already larger than one fifth of the amplitude of $v_\text{s}(t)$. \\

\end{document}